\documentclass[a4paper,11pt]{article}
\usepackage{jinstpub} 
\usepackage{lineno}
\usepackage{subcaption}
\usepackage{siunitx}
\usepackage{nicefrac}

\newcommand{\MeV}[1]{\qty{#1}{\mega\electronvolt}}

\newcommand{\mev}{\mbox{ MeV}}

\title{\boldmath Non-Destructive Beam Monitoring via Secondary Radiation Detection with Ce-Doped Silica Fibers}

\author[a,1]{Alexander Gottstein,\note{Corresponding author. }}
\author[a,b,c]{Pierluigi Casolaro,}
\author[a]{Gaia Dellepiane,}
\author[a]{Lars Eggimann,}
\author[a]{Eva Kasanda,}
\author[a]{Isidre Mateu,}
\author[d]{Samuel Usherovich,}
\author[a,b]{Paola Scampoli,}
\author[d]{Cornelia Hoehr}
\author[a]{and Saverio Braccini}

\affiliation[a]{Laboratory for High Energy Physics (LHEP), Albert Einstein Center for Fundamental Physics (AEC), University of Bern, Switzerland}

\affiliation[b]{Department of Physics ``Ettore Pancini'', University of Napoli Federico II,\\Complesso Univ. Monte S. Angelo, 80126 Napoli, Italy}

\affiliation[c]{INFN Sezione di Napoli, Complesso Univ. Monte S. Angelo, 80126 Napoli, Italy}

\affiliation[d]{Life Sciences, TRIUMF, Vancouver V6T 2A3, British Columbia, Canada}

\emailAdd{alexander.gottstein@unibe.ch}

\abstract{
Non-destructive beam diagnostics are essential for low-energy medical cyclotrons, where even thin interceptive devices can severely degrade beam quality. We investigate an external fiber monitor (EFM) based on Ce-doped silica scintillating fibers that detects secondary radiation generated at existing beamline components of the \qty{18}{\mev} Bern Medical Cyclotron beam transfer line (BTL). Three use cases were studied: (i) beam intensity monitoring around an electrically isolated, water-cooled beam dump; (ii) beam-loss monitoring around a \qty{10}{\milli\meter} collimator under varying the beam focusing; and (iii) by steering a \qty{6.5}{\milli\meter} $\times$ \qty{6.5}{\milli\meter} beam spot on a beam dump. For case (i), the summed EFM signal exhibits a linear dependence on the current on target over nearly three orders of magnitude. In case (ii), a normalized EFM-based beam-loss proxy scales monotonically with an electrical loss proxy across several focusing settings. Furthermore, opposing-fiber signal ratios provide decoupled, monotonic sensitivity to horizontal and vertical beam displacements.
}


\begin{document}
\maketitle
\flushbottom

\section{Introduction}
\label{sec:intro}

Beam diagnostics play a central role in the operation and optimisation of accelerator facilities, such as medical cyclotrons, which typically accelerate protons to a few tens of MeV. In this low-energy regime, reliable monitoring of beam intensity and position is crucial for efficient radioisotope production and for experimental research in medical applications of particle physics. However, at such low energies, destructive beam diagnostics can introduce significant perturbations: even thin interceptive diagnostic devices may alter the transverse beam profile or degrade the energy, making them potentially unsuitable for certain applications, such as cross section studies near reaction thresholds. This motivates the development of non-destructive monitoring systems that can provide reliable information on beam intensity, losses, and position without introducing material into the beam path.

At the Bern medical cyclotron, a commercial \MeV{18} proton accelerator equipped with a~dedicated research beamline, several interceptive systems are in use for characterization tasks, including the UniBEaM fiber scanner~\cite{auger_accelerator_2016}, the Pi2 scintillation-based profiler~\cite{braccini_two-dimensional_2023}, and the Chromoscope optical imaging system~\cite{gottstein_upgraded_2025}.
While these devices have proven highly effective for characterising the beam, they are inherently interceptive and therefore perturb the beam during measurement, or stop it completely. A~complementary strategy is to exploit the secondary radiation, namely neutrons and photons, generated when the proton beam interacts with surrounding materials. 
By detecting radiation produced at the locations where beam–material interactions occur as part of normal operation,~e.g. around collimators or targets, the beam can be monitored without introducing any additional interceptive elements.


In this work, we provide the first systematic assessment of the external fiber monitor (EFM), a non-interceptive diagnostic concept using Ce-doped optical fibers to detect secondary radiation around the beamline. Building on earlier proof-of-principle work~\cite{usherovich_novel_2024}, we evaluate the device in three operational scenarios (beam-intensity monitoring, beam-loss detection, and beam-position sensitivity). Each application is investigated experimentally and presented in the sections that follow.

\section{Materials and Methods}
\label{sec:materials}

The Bern Medical Cyclotron (BMC), located at the Bern University Hospital (Inselspital), is used both for commercial pharmaceutical isotope production and for multidisciplinary research activities. The IBA Cyclone 18/18 HC proton cyclotron is equipped not only with standard liquid target stations for routine isotope production but also with an external beam transfer line (BTL)~\cite{braccini_new_2013}. The BTL ends in a dedicated research bunker that is well shielded from the cyclotron itself. This arrangement provides high accessibility and facilitates the testing of novel beam and radiation instrumentation, such as the EFM investigated in this study. The BTL is equipped with steering magnets and two quadrupole doublets for beam focusing, enabling operation in either flat-beam or focused-beam configurations. All measurements presented in this work were carried out on the BTL. 

\begin{figure}[h]
\centering
\includegraphics[width=0.55\textwidth]{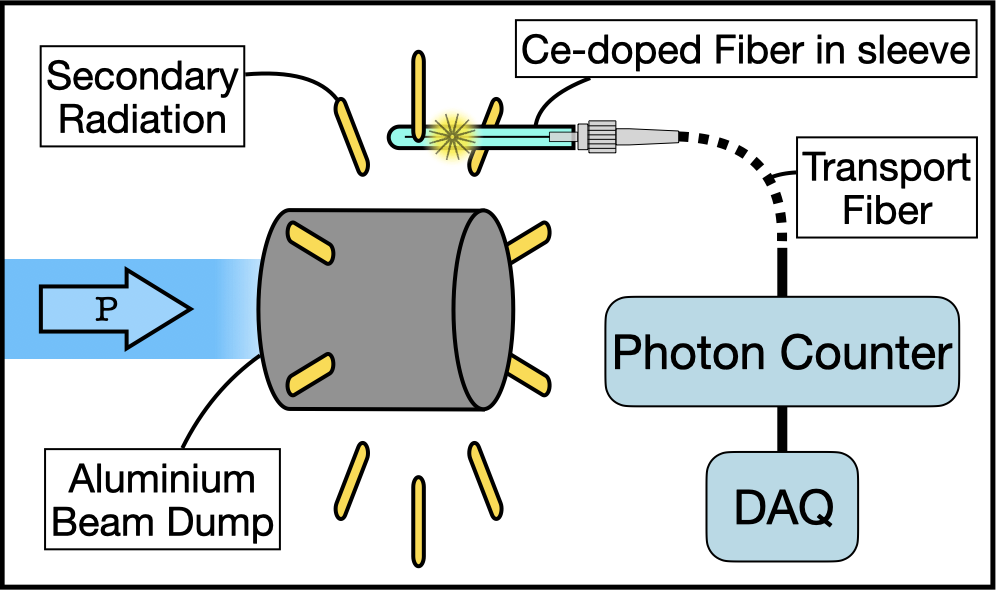}
\caption{Schematic of the measurement setup. The proton beam (in blue) strikes an aluminium beam dump, producing secondary radiation that excites the Ce-doped fibers; the resulting scintillation light is read out with a single-photon counter, connected to a DAQ.}
\label{fig:CollarScheme}
\end{figure}

The EFM consists of Ce-doped silica fibers arranged in an adaptive geometry that allows for mounting around arbitrary beamline components~\cite{usherovich_optical_2022, hoehr_ce-_2020}. The device exploits the fact that secondary radiation (predominantly neutrons and gamma rays) is generated whenever the primary proton beam interacts with material, whether during collimation, through unintended beam losses due to missteering or misfocusing, or upon hitting the intended target. The Ce-doped fibers convert this secondary radiation into scintillation light, which is then transported via optical fibers to the read-out system, where the signal is recorded. The read-out system consists of an IDQ single photon detector~\cite{noauthor_idq_nodate} connected to a Vertilon MCPC618 Eight Channel Pulse Counting System~\cite{noauthor_mcpc618_nodate}. The radiation-sensitive fibers are \qty{3}{\centi\meter} long and are housed in a mating sleeve that shields them from stray light and allows for simple connection to an SMA optical transport fiber. Because the read-out electronics must be kept outside the radiation area, and therefore at a considerable distance from the monitoring fibers, \qty{20}{\meter}-long multimode optical fibers (\qty{200}{\micro\meter} core diameter) are used to transport the signal out of the BTL bunker to the photon counter and data-acquisition system (DAQ) system. The working principle and the general setup are sketched out in Figure~\ref{fig:CollarScheme}.

\subsection{Experimental Setups}
\label{s:BTLsetups}

The EFM was evaluated in three independent configurations corresponding to the three intended use cases: intensity monitoring, beam-loss monitoring, and position monitoring. In all cases the fibers were mounted close to existing beamline components where secondary radiation is produced as part of normal operation. For the linearity (i) and beam-loss monitoring (ii) studies, the fibers were affixed directly to the respective component using adhesive tape. For the beam-position monitoring (iii) measurement, a custom collar-shaped holder was employed to mount four fibers evenly around the beam dump.
For each setup, the transport fibers linking the sensing section to the DAQ were carefully routed away from areas with elevated secondary radiation to prevent unwanted signal pickup (Figure~\ref{fig:watercooled}).

\subsubsection{(i) Linearity Measurement procedure}
\label{sec:linearity}

\begin{figure}[h]
    \centering
    \includegraphics[width=\linewidth]{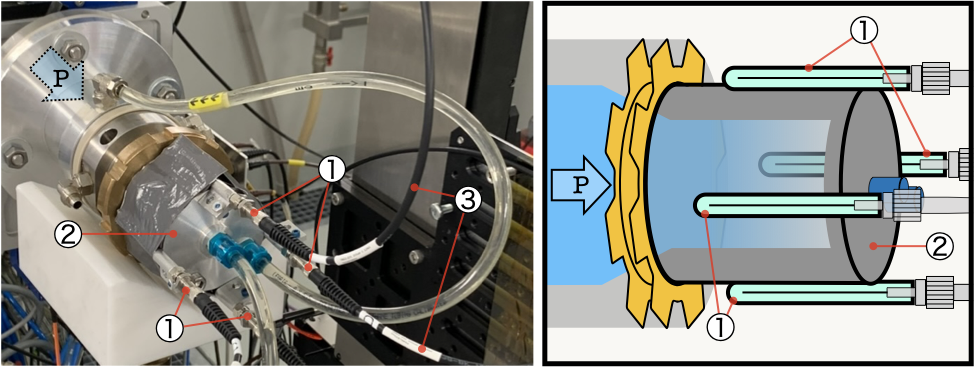}
    \caption{On the left, a photograph of the water-cooled target installed on the BTL is shown. The four EFM sensing fibers (1) are mounted around the beam dump (2). The four transport fibers with the SMA connectors are labelled with (3) in the picture on the left. They are routed away from the beam dump to avoid signal pickup. The proton beam's direction is indicated by the blue arrow. On the right, a simplified schematic of the set-up is shown, showing the internal collimating structure of the target and the axial positions of the fibers around it; schematic not to scale.}
    \label{fig:watercooled}
\end{figure}

To assess the response of the EFM as a function of beam current, a water-cooled aluminium target (Figure~\ref{fig:watercooled}) with a \qty{1}{\centi\meter} circular aperture and an electrically insulated beam dump was used. The beam dump current is measured with a pico-ammeter directly connected to the beam dump. Both the collimator and the beam dump are water-cooled, enabling operation at high beam currents of up to \qty{150}{\micro\ampere}. 

The EFM test fibers were mounted axially around the beam dump, as shown in Figure~\ref{fig:watercooled}. Beam position and stability were monitored with the UniBEaM detector placed upstream. This detector provides a two-dimensional beam profile over a circular area of \qty{3}{\centi\meter} diameter, allowing precise monitoring of the beam position, profile, and stability. 

The EFM response signal was measured over nearly three orders of magnitude in terms of beam current on target. Measurements at each current point were averaged over an interval of approximately 30 seconds. For currents above \qty{40}{\micro\ampere}, the current was steadily increased without stopping the beam between increments to avoid thermal shocks to the target that could physically damage the beam dump. The beam profile and position were not monitored during high‑current runs, as the UniBEaM fibers are not rated for those currents. 

\subsubsection{(ii) Beam Loss Monitoring procedure}
\label{sec:lossmonitoring}

For this measurement, the EFM was installed around a collimator on the BTL to emulate a beamline component that could generate unwanted beam losses (e.g. through losing focus of the beam). Upstream of the collimator, the beam position and shape were characterised with the UniBEaM detector. The beam was then collimated to a circular aperture with a diameter of~\qty{10}{\milli\meter}. The collimator includes a biased electrode ring to suppress the emission of secondary electrons, allowing accurate measurement of the proton current intercepted by the aperture.

\begin{figure}[h]
\centering
\includegraphics[width=\linewidth]{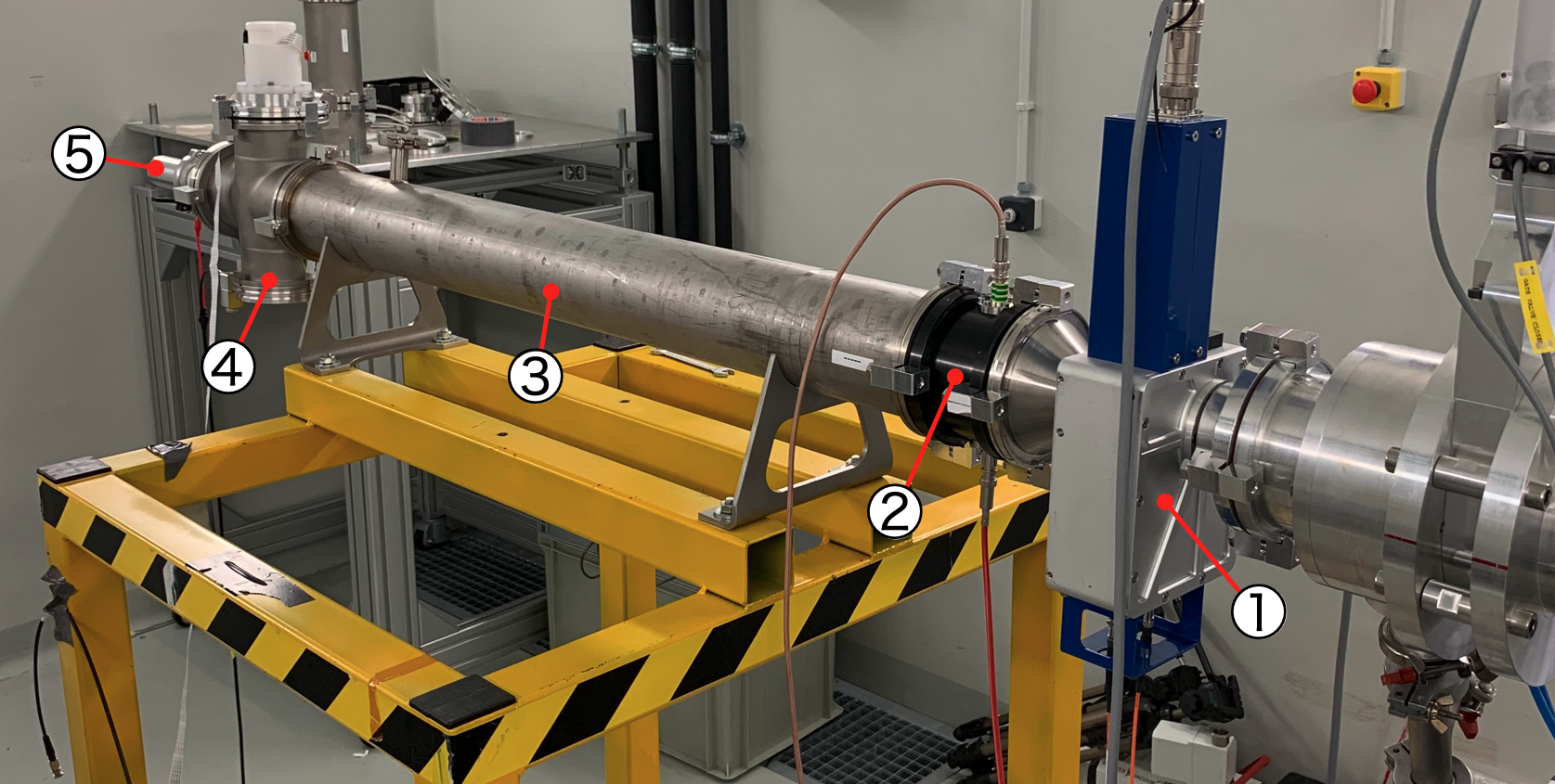}
\caption{The BTL setup used for the \textit{Beam Loss Monitoring} measurement. The FWHM of the proton beam is measured with the UniBEaM (1) right before the beam collimator (2) (\qty{10}{\milli\meter} aperture), where the EFM was installed (not shown in picture). After \qty{1.4}{\meter} of drift space (3) the Pi2 detector (4) is installed before the beam dump (5).}
\label{fig:beamloss}
\end{figure}

Four EFM fibers were mounted around the collimator, each oriented tangentially to the outer circumference of the beam pipe. Downstream of the collimator, the beam was transported over a \qty{1.4}{\meter} drift distance before being stopped in an aluminium beam dump. This distance reduces the contribution of secondary radiation generated at the beam dump to the EFM signal. The beam dump was electrically isolated, enabling the measurement of the total current deposited on target.

During this test, the beam was alternatingly focused and defocused using the BTL's quadrupole magnets located upstream of the UniBEaM. The FWHM of the beam was measured with the UniBEaM upstream of the collimator. Changing the beam focus with the magnetic quadrupoles is not straight-forward, as it often leads to horizontal or vertical steering of the beam. For this reason, an insertable scintillation-based beam profile viewer read out with a CCD camera (referred to as the Pi2~\cite{braccini_two-dimensional_2023}) was installed upstream of the beam dump to verify that the fully collimated beam consistently impinged fully on the beam dump surface. The setup for this measurement is shown in Figure~\ref{fig:beamloss}. 

\subsubsection{(iii) Beam Position Monitoring procedure}
\label{sec:beamsteering}

\begin{figure}[h]
\centering
\subcaptionbox{\label{fig:beamsteering}}{\raisebox{0mm}{\includegraphics[height=5.5cm]{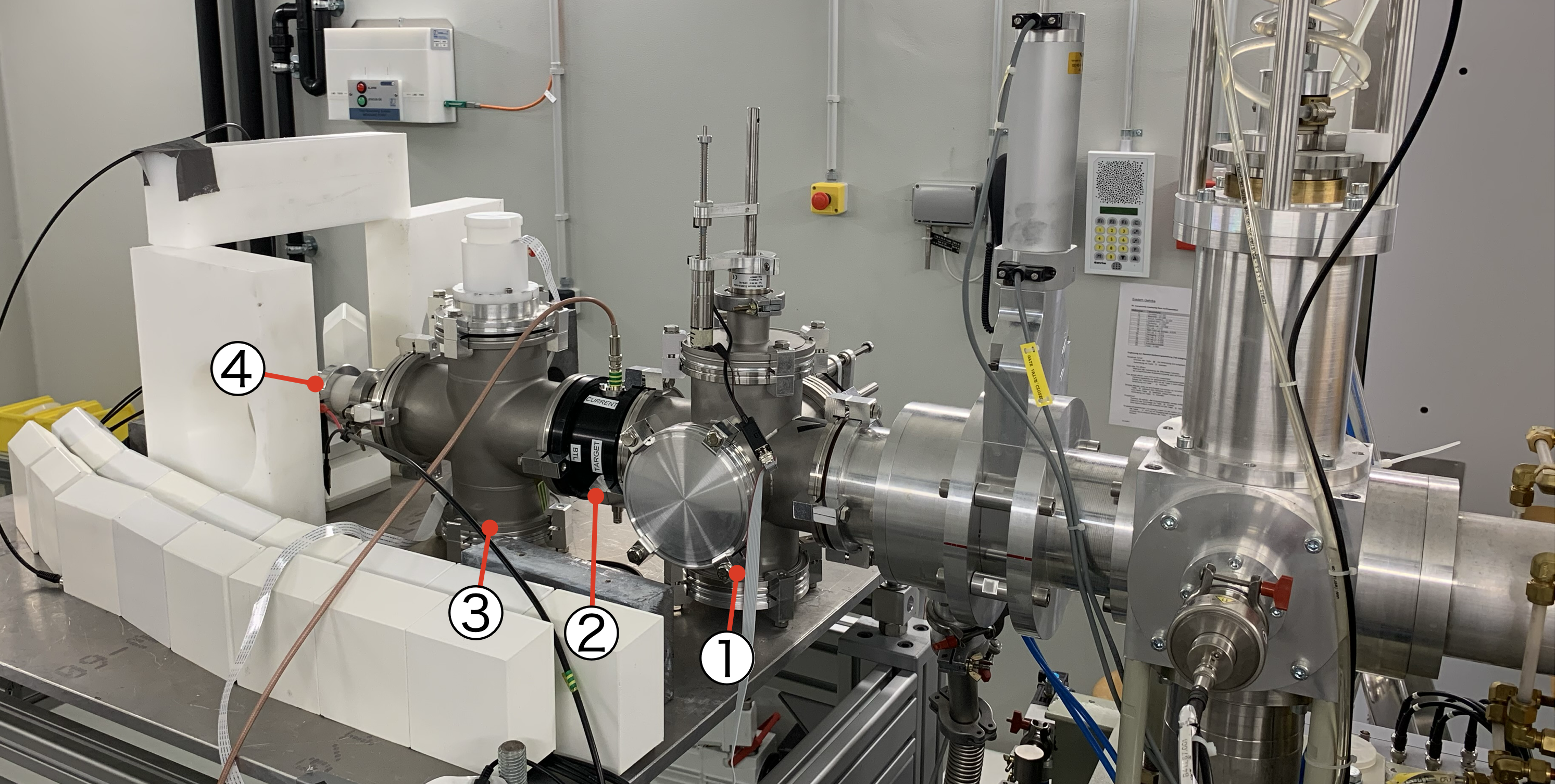}}}
\hfill
\subcaptionbox{\label{fig:collar}}{\raisebox{0mm}{\includegraphics[height=5.5cm]{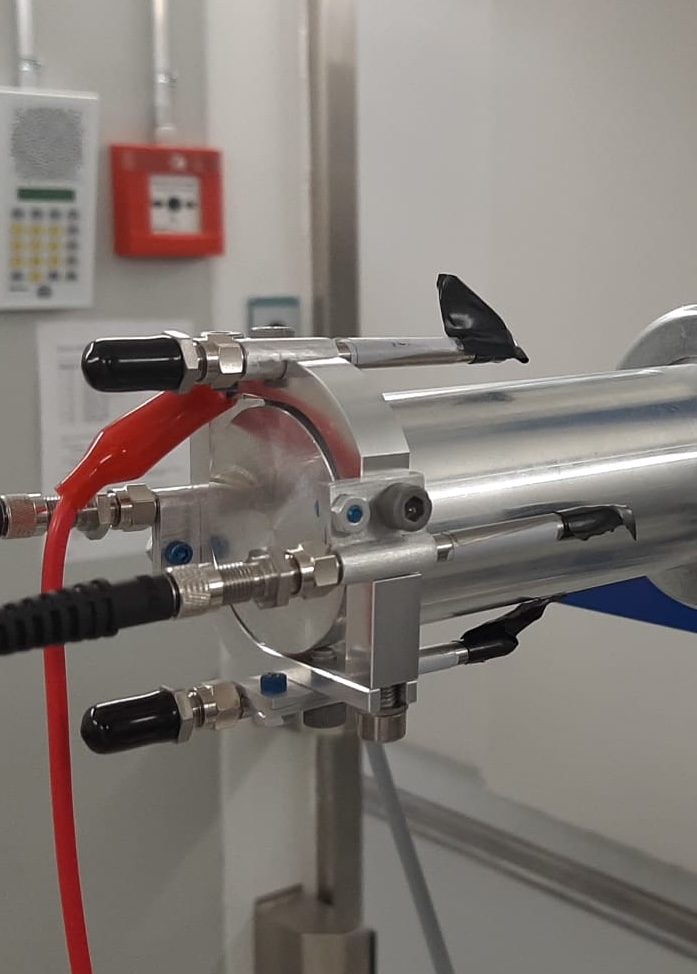}}}
\caption{(a) The BTL setup for the beam steering measurement. The double-slit collimator (1) allows controlled horizontal and vertical steering of a beam spot. An additional collimator (2) is installed to prevent the beam from hitting the beam dump's (4) edge. The Pi2 detector (3) is installed to have an additional reference for the beam spot's position. The EFM fibers are installed around the beam dump at (4). (b) The collar holding the four EFM fibers around the aluminium beam dump.}
\end{figure}

The experimental setup used to assess the performance of the EFM as a beam-position monitor is shown in Figure~\ref{fig:beamsteering}. The EFM fibers were arranged around an aluminium beam dump using a collar-like holder (Figure~\ref{fig:collar}). The beam spot was deliberately steered across the beam dump in both the horizontal and vertical directions. Beam steering was achieved using a movable \textit{double-slit collimator}~\cite{andrey_novel_2023} installed upstream of the drift space. This device provides a square, adjustable aperture whose size and horizontal and vertical position can be controlled, enabling controlled displacement of the beam spot. Compared to dipole-based steering, this approach offers a more direct and stable method, particularly when irradiated with a flat beam-profile, where the extracted beam is less prone to drifts and instabilities compared to when a focused beam is employed. The double-slit collimator, therefore, allows the extraction and controlled steering of a stable square-shaped beam spot, suitable for systematic beam shaping and alignment studies. For this measurement, the aperture was set to a square opening of \qty{6.5}{\milli\meter} $\times$ \qty{6.5}{\milli\meter}. 


Downstream of the double-slit collimator, an additional collimator was installed that had the same size as the beam dump. Its function was purely diagnostic: a signal from this device indicated that the proton beam had reached the geometrical limits of the beam dump. Further downstream, the Pi2 beam profiler was positioned to assess possible inhomogeneities in the extracted beam. In addition, its presence provided supplementary drift space, thereby mitigating the impact of secondary radiation generated at the double-slit collimator on the EFM fibers. The beamline terminated with an electrically insulated aluminium beam dump, enabling direct measurement of the proton current. Four EFM fibers were mounted around the beam dump parallel to the beam direction, with fibers positioned above, below, left, and right of the beam axis. 
\section{Results and Discussion}
\label{sec:results}

The results obtained for the three EFM use cases - (i) linearity monitoring, (ii) beam-loss monitoring, and (iii) beam-position monitoring - are presented and discussed in the following sections.

\subsection{Signal Linearity}

\begin{figure}[htb]
    \centering
    \includegraphics[width=\linewidth]{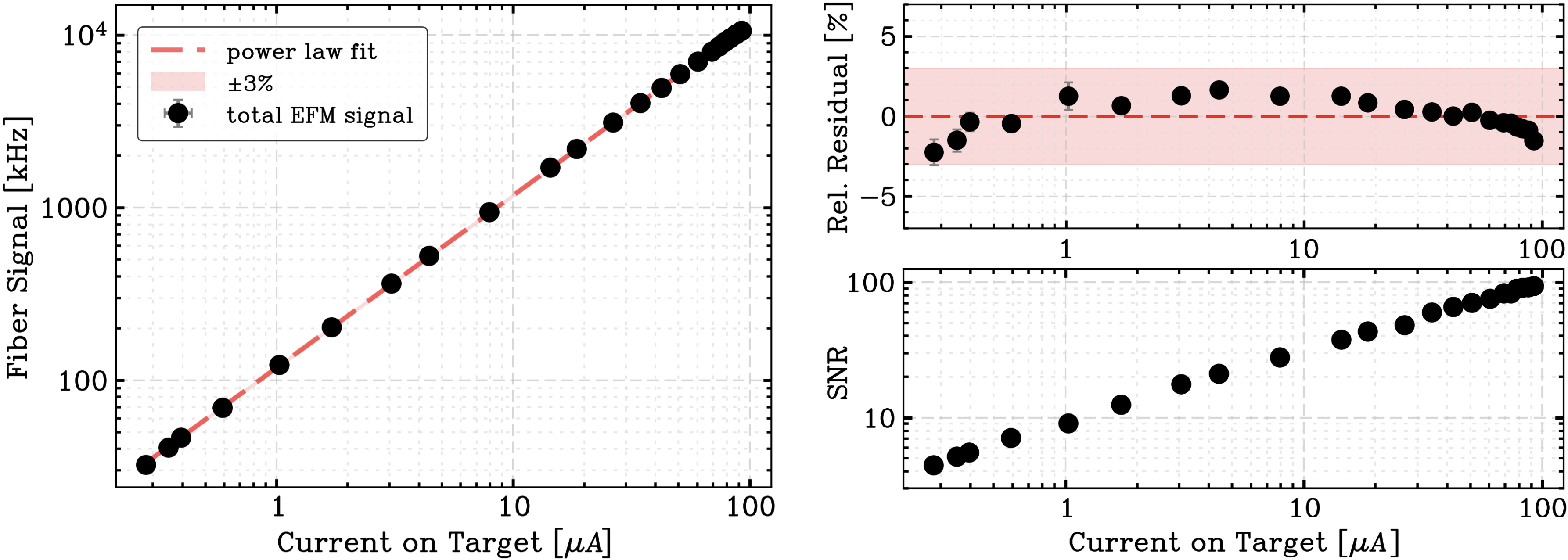}
    \caption{Recorded EFM signal strength is plotted as a function the proton current measured on the water-cooled beam dump. The background subtracted data are fitted with a linear fit on the logarithmic scale (power law fit). 
    The top-right panel shows the relative residuals between the data and the fit, with the $\pm3\%$ interval highlighted in red. The bottom-right panel displays the background-corrected signal-to-noise ratio for each data point.
    }
    \label{fig:linearityplots}
\end{figure}

To assess linearity, the combined count rate from all four EFM fibers was compared with the proton current measured on the water‑cooled beam dump.
A background level measured prior to irradiation was subtracted from all signals. The resulting background-subtracted EFM signal is shown as a function of the target current in Figure~\ref{fig:linearityplots} (left). A least-squares fit with a power‑law model was performed in log–log space (equivalent to a linear fit through the origin in linear space). The corresponding relative residuals, defined as the difference between data and fit, normalised to the fit, are shown in the top‑right panel. The residuals exhibit a small systematic drift across the measured current range but remain within $\pm3\%$, indicating a linear response over the explored range.


The deviation observed at low currents is attributed to known beam instabilities. The UniBEaM detector, used for alignment and focusing of the initial beam, cannot withstand high currents and was switched off after the calibration. Following this step, beam stability was monitored via the beam dump current. As this method provides no direct beam-position information, small positional drifts at the target cannot be excluded. Because the four fibers have different overall sensitivities, such shifts can change the summed EFM signal and produce the drift seen in the low current regime.\newline
The drift in the high current regime is attributed to the activation of the target induced by the proton beam. During the irradiation of aluminium with protons, the relatively short-lived isotope $^{27}$Si is produced~($\,T_{\nicefrac{1}{2}} =$ \qty{4.15}{\second}). To avoid thermal damage to the beam dump, measurements at currents above \qty{20}{\micro \ampere} were limited to a duration of 20 seconds. If the irradiation time is shorter than the time required to reach an equilibrium between the production and decay of $^{27}$Si (approximately \qty{25}{\second}), a reduced signal yield is expected. Calculations using \cite{koning_medical_2020} show that increasing the irradiation duration by about \qty{5}{\second} would increase the contribution from the target activation by about $8\%$, which could account for the $2\%$ residual difference of the overall signal at the high currents.

Error bars on the plots reflect only the statistical uncertainty of the averaged EFM signal, while the systematic uncertainties described above are not included. The bottom‑right panel of Figure~\ref{fig:linearityplots} shows the background‑subtracted signal‑to‑noise ratio (SNR) as a function of target current: the SNR falls to a value of about 4.4 at the lowest measured currents and increases roughly linearly to a value of about 94 at the maximum measured current of \qty{92}{\micro\ampere}. The reduced SNR at low currents explains the larger residual uncertainties there and sets a practical lower limit for quantitative linear response.

\begin{figure}[h]
\centering
\subcaptionbox{\label{fig:sub11}}{\raisebox{0.75mm}{\includegraphics[height=5.2cm]{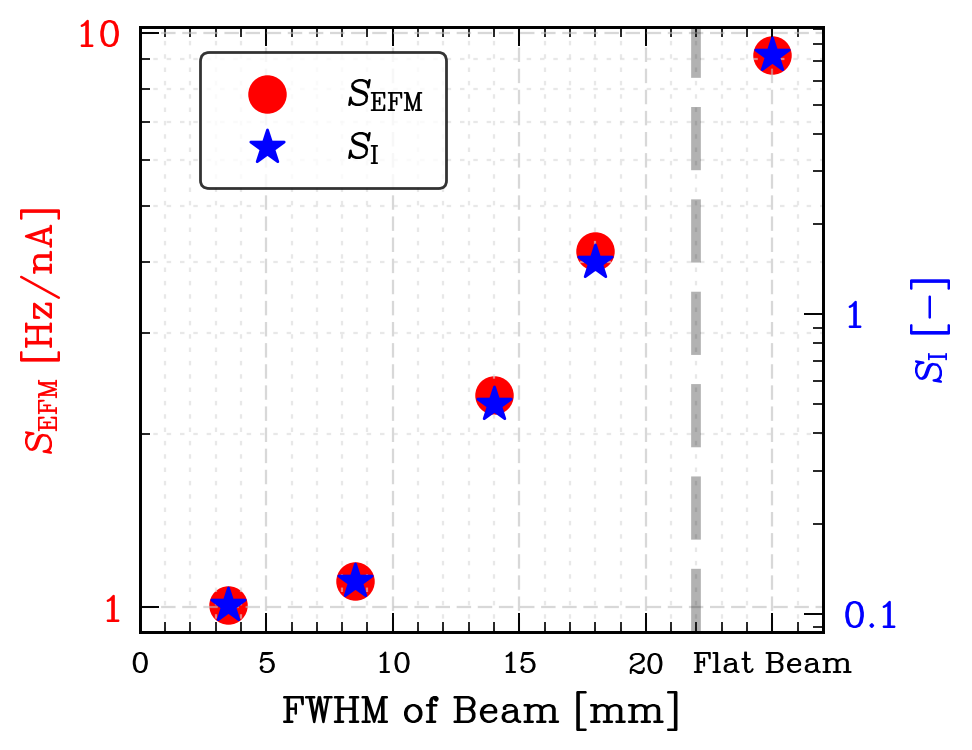}}}
\hfill
\subcaptionbox{\label{fig:sub22}}{\raisebox{0mm}{\includegraphics[height=5.22cm]{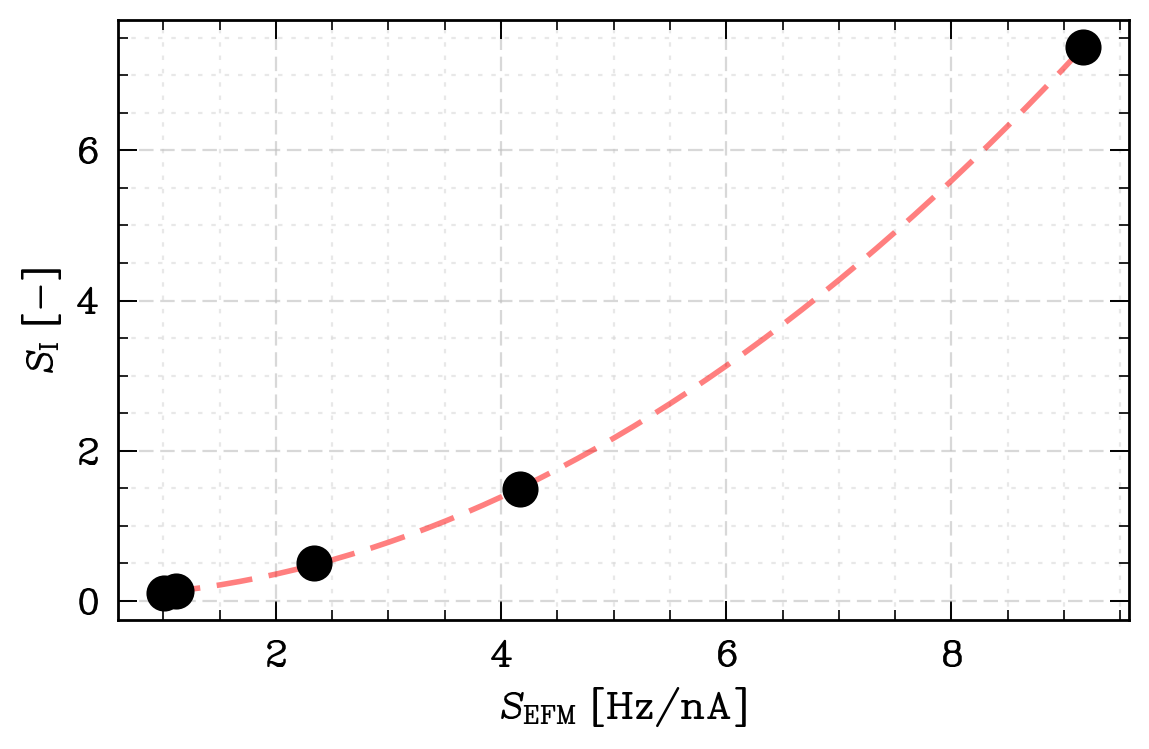}}}
\caption{
In panel (a)  the proxy signals defined in Eqs.~(\ref{eq:SI}) and~(\ref{eq:SEFM}) as a function of the FWHM of the collimated proton beam. Panel (b) shows the beam-loss factor  $S_\text{I}$ as a function of $S_\text{EFM}$ together with a quadratic fit to their correlation. The fit provides an empirical calibration curve. Statistical uncertainties are negligible and therefore not visible in either panel.
}
\label{fig:beamloss_results}
\end{figure}

\subsection{Beam Loss Monitoring}

 By varying the beam focus from a tightly focused spot (FWHM = \qty{3.5}{\milli\meter}) to a quasi-flat beam, we compared the scaling behaviour of the two `beam loss factors' $S_{I}$ and $S_\text{EFM}$, defined as

\noindent
\begin{minipage}[t]{0.34\textwidth}
\begin{equation}\label{eq:SI}
  S_{I} = \frac{I_\text{coll}}{I_\text{tot}},
\end{equation}
\end{minipage}
\begin{minipage}[t]{0.23\textwidth}
  \centering \vspace{3mm}\qquad \qquad \quad and \! 
\end{minipage}
\begin{minipage}[t]{0.4\textwidth}
\begin{equation}\label{eq:SEFM}
  S_\text{EFM} = \frac{C_\text{EFM}}{I_\text{tot}},
\end{equation}
\end{minipage}
\vspace{3mm}

in different scenarios where beam is lost on a beam line component, i.e. the beam collimator. In the equations above, $I_\text{coll}$ is the proton current measured on the collimator, and $I_\text{tot}$ is the sum of the currents measured on the collimator and beam dump, i.e.\ the total current of the system. $C_\text{EFM}$ denotes the total EFM signal count rate. Thus, the proxies $S_{I}$ and $S_\text{EFM}$ represent the collimator and EFM signals, respectively, normalised to the total proton current extracted from the cyclotron, and thus act as indirect measures of the beam intensity. For this measurement, $I_\text{tot}$ was kept at approximately \qty{1}{\micro \ampere} throughout all irradiations.

Figure~\ref{fig:sub11} shows $S_I$ and $S_\text{EFM}$ for five beam FWHM values, plotted on separate y‑axes. Both proxies increase monotonically with increasing FWHM, allowing a direct calibration between them. 
Figure~\ref{fig:sub22} shows $S_\text{EFM}$ as a function of $S_{I}$, together with a quadratic fit. This relation is used as an empirical calibration curve to convert EFM measurements into an estimate of beam loss at the collimator associated with variations in beam focusing. This calibration is strongly dependent on the collimator geometry, in particular the collimator aperture, and it is not universally applicable. 

\subsection{Beam Position Monitoring}

Signals from the four EFM fibers positioned around the beam dump (left, right, top, and bottom) were recorded while the $6.5~\text{mm} \times 6.5~\text{mm}$ beam spot was moved across the beam dump. Two independent scans were performed: in the first, the beam was steered horizontally from left to right on the beam dump; in the second, it was steered vertically from top to bottom on the beam dump. 

\begin{figure}[h]
\centering
\vspace{5mm}
\subcaptionbox{\label{fig:LtoR}}{\includegraphics[width=7.3cm]{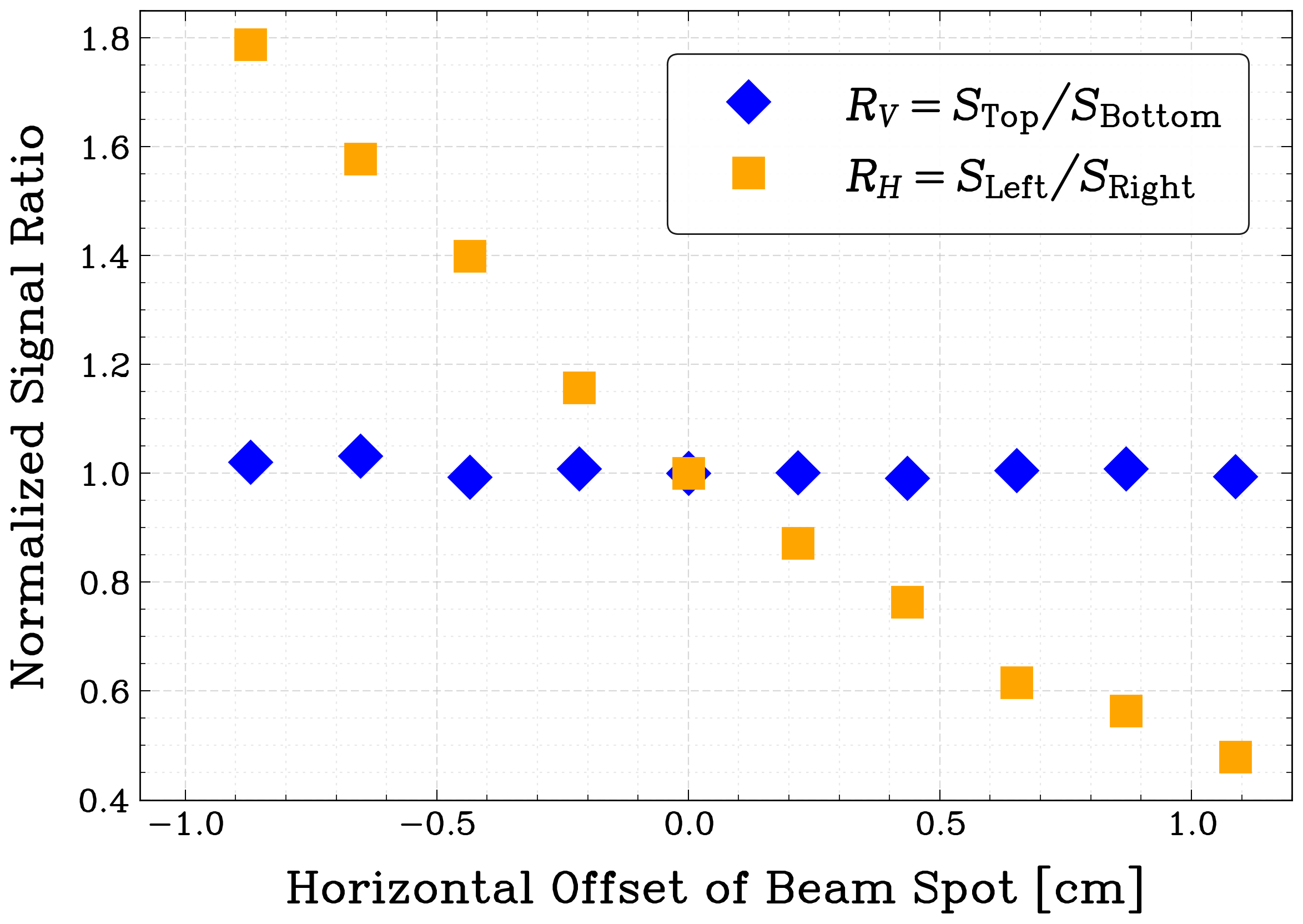}}\hfill
\subcaptionbox{\label{fig:TtoB}}{\includegraphics[width=7.3cm]{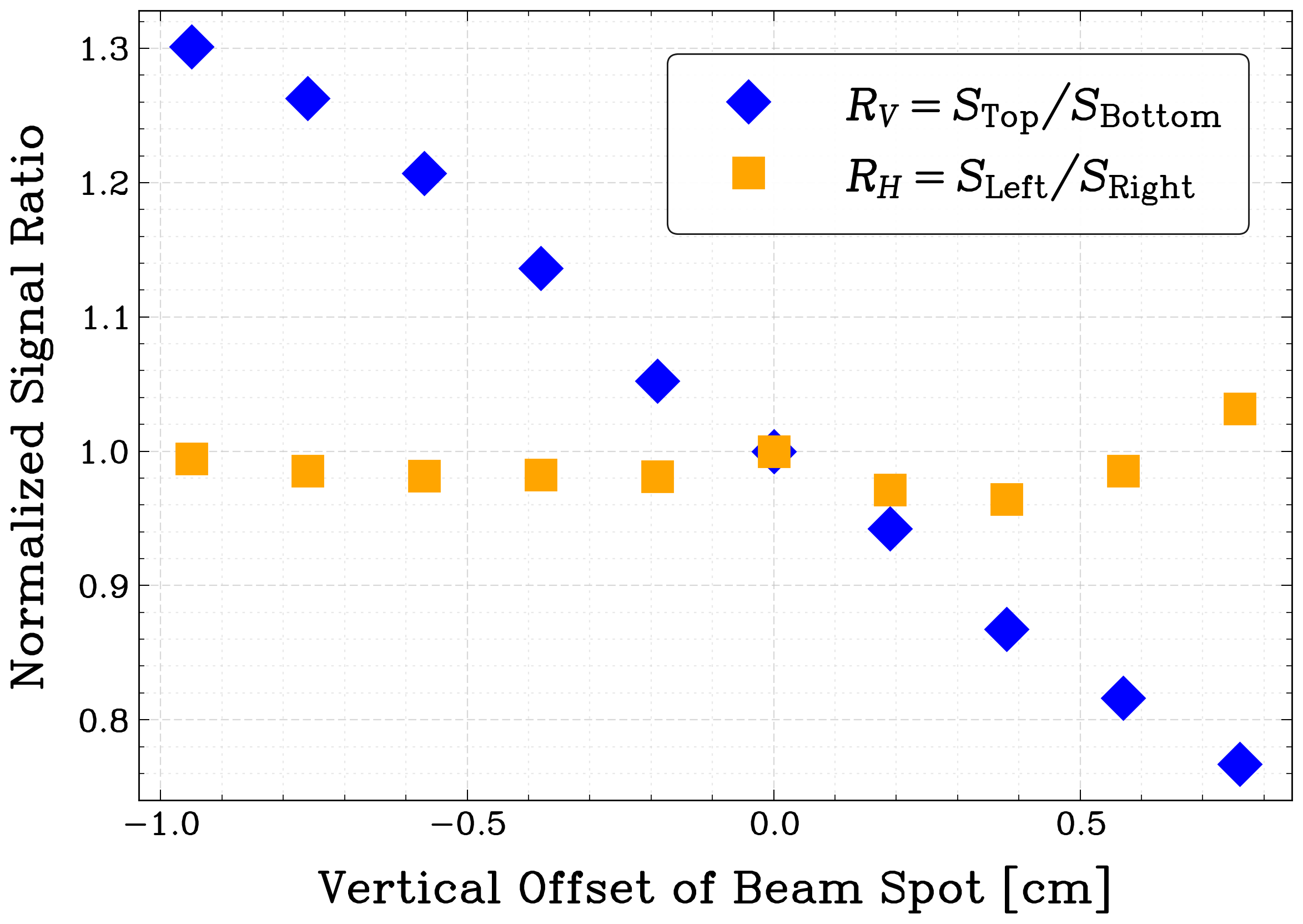}}
\caption{Signal ratios $R_H$ and $R_V$ are recorded during horizontal (a) and vertical (b) steering of the beam across the beam dump. In (a), negative offset values correspond to beam positions shifted to the left, while in (b), negative offsets correspond to beam positions closer to the top fiber.}
\label{fig:beamsteering2}
\end{figure}

Figure~\ref{fig:beamsteering2} presents the resulting horizontal ratio $R_H = \nicefrac{S_\text{Left}}{S_\text{Right}}$ of the signals $S$ of the left and the right fiber, and its vertical counterpart $R_V = \nicefrac{S_\text{Top}}{S_\text{Bot. }}$. A~clear dependence of these ratios on the beam position is observed. During the left-to-right steering sequence, the horizontal ratio $R_H$ varies steadily with the lateral position of the beam, while the vertical ratio $R_V$ remains essentially constant. Conversely, during the top-to-bottom steering sequence, only the vertical ratio $R_V$ changes, whereas the horizontal ratio $R_H$ remains stable. These results demonstrate that the relative signal between opposing fiber pairs provides a direct and decoupled measure of the beam position along each axis. The SNR for each individual fiber was approximately $\sim$80 throughout the measurement. 

\newpage In the top-to-bottom steering data shown in Figure~\ref{fig:TtoB}, the horizontal ratio $R_H$ does not remain constant as the beam approaches the lower region of the beam dump. The observed features are consistent with a beam drift occurring during the measurement. If the profile of the extracted beam from the cyclotron is flattened using the available focusing magnets, small thermally induced variations in the cyclotron's magnetic field can lead to a slow horizontal displacements of the extracted beam or to development of gradients across the flat beam profile. The observed behaviour is consistent with and could be explained by this mechanism.

\section{Conclusions}
\label{s:conclusions}

A non-destructive external fiber monitor (EFM) based on Ce-doped silica fibers has been evaluated for beam diagnostics through the detection of secondary radiation from beam–material interactions. The system was tested on the Bern Medical Cyclotron beam transfer line in three configurations: (i) intensity monitoring, (ii) beam-loss monitoring, and (iii) position monitoring. 

In the intensity-monitoring configuration, the summed EFM signal exhibits a linear dependence on beam current over three orders of magnitude. A systematic drift in the residuals is observed, which are attributed to uncertainties arising from unstable beam parameters, i.e. transverse profile and position stability. 

For the beam‑loss monitoring, the EFM signal measured around the collimator scales monotonically with the electrical loss proxy $S_\text{I}$ (derived from the current measurement), as the beam FWHM is varied. This monotonic relationship indicates that, in similar configurations, a calibration curve can be established that allows beam losses at the collimator to be inferred directly from the EFM signal.\newline
Employing a spatial array of fibers would make it possible to distinguish losses arising from changes in beam focusing from those caused by beam drifts. This is supported by our third experiment, in which an array of four fibers was used as a beam‑position monitor, with opposing‑fiber signal ratios providing decoupled, monotonic sensitivity to horizontal and vertical steering. 

The EFM is non-interceptive and can be easily retrofitted around existing components with remote readout. It can complement interceptive diagnostics (e.g. SEMs, wire grids, collimators, faraday cups, or scintillator screens), enabling continuous monitoring, e.g. during production of radioisotopes, and providing additional observables (losses near apertures, position trends) without additionally perturbing the beam.

The EFM response, however, is geometry- and material-dependent, requiring site-specific calibration. The same applies for the signal-to-noise ratio of the measured signal. Furthermore, transport-fiber pickup should be mitigated either by careful routing away from radiation hotspots or by differential readout using a second reference transport fiber for background subtraction. Another limitation arises from the activation of targets which produces an additional signal picked up by the fibers. This does not pose a concern for the results shown in this work except where explicitly noted (i.e. during high-current irradiations in measurement (i)), as the irradiations were performed 
over a sufficiently long  time for an equilibrium between the production and decay of aluminium isotopes to be established. If the secondary radiation signal of an aluminium target is measured using beams with pulse durations on the order of the half-life of~$^{27}$Si~($\,T_{\nicefrac{1}{2}} =$ \qty{4.15}{\second}), this effect must be taken into account. For targets composed of materials other than aluminium, different radionuclides are produced, and the corresponding critical time scales will change accordingly.

The EFM potentially provides a practical and sensitive non-destructive diagnostic for beam intensity, losses, and steering at medical-cyclotron energies and, in principle, also at higher energies. With site-specific calibration and appropriate systematic controls, it can be integrated into routine operations as a complementary beam-quality monitor.

\section*{Outlook}

To enhance the EFM's performance, we plan to replace the Ce-doped silica fibers with a higher–light-yield scintillator like GAGG:Ce, to improve the signal-to-noise ratio and extend the usable dynamic range, especially at low currents. The increased light yield would also allow a reduction of reducing the scintillator's sensitive volume, 
thereby localising the pickup of secondary radiation and improving the spatial precision of the measurement. This is potentially useful for beam position monitoring at length scales on the order of the current fiber size. Moreover, a higher signal level would reduce the relative impact of unwanted light collected by the transport fibers and thus enhance the reliability of the EFM. In preliminary tests under comparable conditions with a $5\times5\times5$~\qty{}{\milli\meter\cubed} GAGG:Ce crystal, the measured signal exceeded that of the Ce-doped fibers by approximately an order of magnitude. Ongoing work focuses on optimising material choice, sensor geometry, and readout.
Additional improvements could come from the implementation of signal discrimination between neutron- and gamma-induced contributions, which constitute the dominant components of secondary radiation. By separating these components, it may become possible to quantify the fraction of the signal arising from target activation. Achieving this would require the use of scintillators with appropriate particle-discrimination capabilities and sufficiently high signal yield. Also using two different scintillators with different neutron- and gamma-sensitivities could be an option.
\acknowledgments

This research was partially funded by the Swiss National Science Foundation (SNSF). Grants: IZURZ2\_224901 and CRSII5\_180352. We are grateful for the contributions of the LHEP engineering and technical staff. \hfill \break 


\noindent \textbf{Data Availability:} Data will be made available on request.

\newpage
\bibliographystyle{JHEP}
\bibliography{references}

@misc{noauthor_idq_nodate,
	type = {Product {Sheet}},
	title = {{IDQ} {ID100} {Brochure}},
	url = {https://marketing.idquantique.com/acton/attachment/11868/f-0236/1/-/-/-/-/ID100_Brochure.pdf},
	language = {en},
	urldate = {2025-04-03},
	publisher = {ID Quantique SA},
	note = {URL: https://marketing.idquantique.com/acton/attachment/11868/f-0236/1/-/-/-/-/ID100\_Brochure.pdf (Accessed: 03.04.2025)},
}

@misc{noauthor_mcpc618_nodate,
	type = {Product {Sheet}},
	title = {{MCPC618} {Eight} {Channel} {Pulse} {Counting} {System}},
	shorttitle = {Vertilon {MCPC618}},
	url = {https://vertilon.com/pdf/PS2720.pdf},
	language = {en},
	urldate = {2026-01-01},
	publisher = {Vertilon Corporation},
}

@phdthesis{andrey_novel_2023,
	address = {Bern},
	type = {Bachelor {Thesis}},
	title = {A novel double-slit collimator for the {Bern} medical cyclotron},
	language = {en},
	school = {University of Bern},
	author = {Andrey, Alan},
	year = {2023},
}

@incollection{koning_medical_2020,
	title = {The medical isotope browser},
	issn = {0074-1884},
	url = {https://inis.iaea.org/records/32fc0-sp210},
	language = {English},
	urldate = {2025-12-23},
	booktitle = {Trends in {Radiopharmaceuticals} ({ISTR}-2019). {Proceedings} of an {International} {Symposium}. {Programme} and {Abstracts}},
	author = {Koning, Arjan and Verpelli, Marco},
	month = nov,
	year = {2020},
	pages = {116--116},
}

@inproceedings{gottstein_upgraded_2025,
	address = {Liverpool, UK},
	title = {Upgraded beam profile monitoring using {Chromox} and fiber optic imaging for high-radiation environments},
	url = {https://meow.elettra.eu/90/},
	doi = {10.18429/JACoW-IBIC2025-TUPMO16},
	language = {en},
	booktitle = {Proceedings of the 14th {International} {Beam} {Conference}},
	author = {Gottstein, Alexander and Kasanda, Eva and Mateu, Isidre and Eggimann, Lars and Braccini, Saverio},
	month = sep,
	year = {2025},
}

@article{usherovich_novel_2024,
	title = {A novel fiber-optic beam monitor},
	volume = {2687},
	issn = {1742-6588, 1742-6596},
	url = {https://iopscience.iop.org/article/10.1088/1742-6596/2687/7/072003},
	doi = {10.1088/1742-6596/2687/7/072003},
	language = {en},
	number = {7},
	urldate = {2025-08-28},
	journal = {Journal of Physics: Conference Series},
	author = {Usherovich, Samuel and Casolaro, Pierluigi and Gottstein, Alexander and Mateu, Isidre and Dehnel, Morgan and Braccini, Saverio and Hoehr, Cornelia},
	month = jan,
	year = {2024},
	pages = {072003},
}

@article{braccini_two-dimensional_2023,
	title = {A {Two}-{Dimensional} {Non}-{Destructive} {Beam} {Monitoring} {Detector} for {Ion} {Beams}},
	volume = {13},
	copyright = {https://creativecommons.org/licenses/by/4.0/},
	issn = {2076-3417},
	url = {https://www.mdpi.com/2076-3417/13/6/3657},
	doi = {10.3390/app13063657},
	language = {en},
	number = {6},
	urldate = {2025-04-29},
	journal = {Applied Sciences},
	author = {Braccini, Saverio and Carzaniga, Tommaso Stefano and Casolaro, Pierluigi and Dellepiane, Gaia and Franconi, Laura and Mateu, Isidre and Scampoli, Paola and Schmid, Matthias},
	month = mar,
	year = {2023},
	pages = {3657},
}

@article{hoehr_ce-_2020,
	title = {Ce- and {B}-{Doped} {Silica} {Fibers} for {Monitoring} {Low}-{Energy} {Proton} {Beams} on a {Medical} {Cyclotron}},
	volume = {10},
	copyright = {https://creativecommons.org/licenses/by/4.0/},
	issn = {2076-3417},
	url = {https://www.mdpi.com/2076-3417/10/13/4488},
	doi = {10.3390/app10134488},
	language = {en},
	number = {13},
	urldate = {2025-04-03},
	journal = {Applied Sciences},
	author = {Hoehr, Cornelia and Hanna, Matthew and Zeisler, Stefan and Penner, Crystal and Stokely, Matthew and Dehnel, Morgan},
	month = jun,
	year = {2020},
	pages = {4488},
}

@article{auger_accelerator_2016,
	title = {Accelerator and detector physics at the {Bern} medical cyclotron and its beam transport line},
	volume = {61},
	copyright = {http://creativecommons.org/licenses/by-nc-nd/3.0},
	issn = {0029-5922},
	url = {https://www.sciendo.com/article/10.1515/nuka-2016-0009},
	doi = {10.1515/nuka-2016-0009},
	language = {en},
	number = {1},
	urldate = {2025-02-05},
	journal = {Nukleonika},
	author = {Auger, Martin and Braccini, Saverio and Ereditato, Antonio and Häberli, Marcel and Kirillova, Elena and Nesteruk, Konrad P. and Scampoli, Paola},
	month = mar,
	year = {2016},
	pages = {11--14},
}

@article{usherovich_optical_2022,
	title = {Optical fibre array detector to monitor irradiations for medical radioisotope production},
	volume = {2374},
	issn = {1742-6588, 1742-6596},
	url = {https://iopscience.iop.org/article/10.1088/1742-6596/2374/1/012182},
	doi = {10.1088/1742-6596/2374/1/012182},
	language = {en},
	number = {1},
	urldate = {2022-12-20},
	journal = {Journal of Physics: Conference Series},
	author = {Usherovich, Samuel and Penner, Crystal and Hodgson, Geoff and Thoeng, Edward and Dehnel, Morgan and Hoehr, Cornelia},
	month = nov,
	year = {2022},
	pages = {012182},
}

@article{braccini_new_2013,
	title = {The new bern {PET} cyclotron, its research beam line, and the development of an innovative beam monitor detector},
	volume = {1525},
	issn = {0094-243X},
	url = {https://aip.scitation.org/doi/abs/10.1063/1.4802308},
	doi = {10.1063/1.4802308},
	number = {1},
	urldate = {2022-12-19},
	journal = {AIP Conference Proceedings},
	publisher = {American Institute of Physics},
	author = {Braccini, Saverio},
	month = apr,
	year = {2013},
	pages = {144--150},
}

\end{document}